\documentclass[a4paper,11pt]{article}
\pdfoutput=1 

\usepackage{jheppub} 

\usepackage[T1]{fontenc} 
\usepackage{graphicx}
\usepackage{amsthm}
\usepackage{amsmath}
\usepackage{graphicx,subfigure}
\usepackage{dcolumn}
\usepackage{bm}
\usepackage{slashed}
\usepackage{calligra}

\title{\boldmath Another relation among the neutrino mass-squared differences?}

 \author{I. Alikhanov}
 \affiliation{North-Caucasus Center for Mathematical Research, North-Caucasus Federal University,\\
Stavropol 355017, Russian Federation}

\emailAdd{ialspbu@gmail.com}

\abstract{Determining the absolute neutrino mass scale remains a compelling challenge in particle physics. Establishing precise mathematical connections among the neutrino masses could significantly facilitate  this task and shed additional light on the underlying mass-generation mechanism. We highlight the possible existence of a simple algebraic relation between the mass-squared differences, $$\frac{\sqrt{\Delta m^2_{31}}+\sqrt{\Delta m^2_{21}}}{\sqrt{\Delta m^2_{31}}-\sqrt{\Delta m^2_{21}}}=\sqrt{2},$$ which agrees remarkably well with the best-fit values of recent global analyses of neutrino oscillation data. We then investigate the implications of adopting this relation and how it decreases the number of independent parameters in the neutrino sector. We demonstrate that there are alternative representations well known in quark and charged-lepton physics. Furthermore, we place a tight constraint on the solar neutrino mass splitting, \mbox{$\Delta m^2_{21}=(7.445 \pm 0.062) \times 10^{-5} \text{ eV}^2$}, and also obtain  $\Delta m^2_{31}=(2.548 \pm 0.041) \times 10^{-3} \text{ eV}^2$. Next-generation  reactor and accelerator neutrino experiments are expected to significantly reduce systematic uncertainties. In particular, the Jiangmen Underground Neutrino Observatory (JUNO) will measure these parameters independently of each other at the sub-percent level, thereby providing a stringent and direct test of our results.}

\begin{document} 
\maketitle
\flushbottom

\section{Introduction}
The standard three-neutrino paradigm is based on the principle that three neutrino flavor eigenstates ($\nu_e$,~$\nu_\mu$,~$\nu_\tau$) are mixed with three neutrino mass eigenstates ($\nu_1$, $\nu_2$, $\nu_3$) of masses $m_1$, $m_2$, $m_3$. This mixing is described by the unitary $3\times3$ Pontecorvo--Maki--Nakagawa--Sakata matrix $U_{\ell i}$~\cite{ParticleDataGroup:2024cfk}, typically parametrized by three angles~($\theta_{12}$, $\theta_{23}$, $\theta_{13}$) and a CP-violating phase~($\delta_{\text{CP}}$).

Experiments using solar~\cite{SNO:2002tuh,Super-Kamiokande:2001ljr}, atmospheric~\cite{Super-Kamiokande:1998kpq,IceCube:2014flw}, accelerator~\cite{K2K:2004iot,MINOS:2008kxu,T2K:2011ypd,OPERA:2015wbl}, and reactor~\cite{KamLAND:2002uet,DoubleChooz:2011ymz,DayaBay:2012fng,RENO:2012mkc} neutrino sources have confirmed the existence of oscillations—the transitions between the flavor eigenstates as neutrinos propagate.  Because the standard oscillation mechanism requires nonzero masses, these observations also imply that at least two of the three neutrino states are massive. However, oscillation experiments measure  the neutrino mass-squared differences rather than  the absolute masses themselves.
Determining the latter remains a fundamental yet unsolved problem in particle physics. The Standard Model does not predict the exact neutrino masses. The mass ordering—normal ($m_1<m_2<m_3$) or inverted ($m_3<m_1<m_2$)—has not yet been definitively established, although current data mildly favor normal ordering~\cite{T2K:2023smv,T2K:2025yoy,NOvA:2025tmb, Esteban:2026phq}. However, the first joint analysis of the data sets reported in~\cite{T2K:2023smv,T2K:2025yoy,NOvA:2025tmb} found no strong preference for either ordering~\cite{T2K:2025wet}. The  operating JUNO detector~\cite{JUNO:2022mxj,JUNO:2025gmd} and upcoming projects such as DUNE~\cite{DUNE:2020lwj} and Hyper-Kamiokande~\cite{Hyper-Kamiokande:2018ofw} are designed to determine the neutrino mass ordering with high statistical significance (3--$5\sigma$).

There are three mass-squared differences, $\Delta m^2_{21}=m^2_2-m^2_1$, $\Delta m^2_{31}=m^2_3-m^2_1$, and \mbox{$\Delta m^2_{32}=m^2_3-m^2_2$}, of which two are regarded as independent~\cite{Denton:2025jkt}. For instance, the JUNO collaboration has reported the most precise measurement of \(\Delta m^2_{21}\) to date, \mbox{\((7.50 \pm 0.12) \times 10^{-5}\)~eV\({}^{2}\)}~\cite{JUNO:2025gmd}. The NuFIT-6.1 global analysis of neutrino oscillation data, including the JUNO latest results, yields \mbox{$\Delta m^2_{31}=(2.529\pm0.021)\times10^{-3}~\text{eV}^2$} for normal ordering\footnote{Throughout this work, we assume normal  ordering ($\Delta m^2_{31}>0$).}~\cite{Esteban:2026phq}. Both parameters are expected to be measured independently at a sub-percent level in the near future, particularly by JUNO~\cite{JUNO:2022mxj}.  In this regard, it becomes important to take into account sub-leading corrections, such as matter effects~\cite{Khan:2019doq}. Nonetheless, these observables and their inherent  consistency condition, \mbox{$\Delta m^2_{31}-\Delta m^2_{21}=\Delta m^2_{32}$}, are insufficient to determine the absolute neutrino masses;  the number of unknowns exceeds the number of equations. Therefore, additional precise mathematical relations among the neutrino mass-squared differences could aid  this determination. 
They could provide useful calculational tools as well as insights into the underlying mass-generation mechanism. 
The fact that neutrinos  are at least six orders of magnitude lighter than the charged leptons points toward physics beyond the Standard Model. 

Information on the absolute neutrino mass scale is obtained primarily from direct measurements, such as the KATRIN experiment~\cite{Drexlin:2026zam}, and from cosmological observations~\cite{Dolgov:2002wy}. KATRIN probes the endpoint region of the tritium beta spectrum and provides a model-independent upper bound on the effective "electron neutrino mass",~$m_{\nu_e}=$~$\sqrt{\sum_i|U_{ei}|^2m^2_i}$. Cosmology—while model-dependent—constrains the sum of neutrino masses, $\sum m_\nu=m_1+m_2+m_3$. Both approaches have significantly narrowed down the unexplored parameter space. For example, the KATRIN experiment has recently set the limit~\mbox{$m_{\nu_e}<0.45$~eV} at 90\% confidence level~(C.L.)~\cite{KATRIN:2024cdt} and is expected to reach its final target sensitivity of less than~0.3~eV~\cite{Drexlin:2026zam}. The DESI collaboration reports a bound of \mbox{$\sum m_\nu<0.072$~eV}~(95\%~C.L.)~\cite{DESI:2024mwx}, which, however, can be relaxed to $\sum m_\nu<0.11$~eV under alternative analysis methods or data-set combinations~\cite{Naredo-Tuero:2024sgf,Gorbunov:2026sly}. Another quantity of central interest is the effective Majorana mass $m_{\beta\beta}=|\sum_iU_{ei}^2m_i|$, probed in searches for neutrinoless double-beta ($0\nu\beta\beta$) decay. In contrast to~$m_{\nu_e}$, $m_{\beta\beta}$ depends on unknown  phases~\cite{DellOro:2016tmg,Gomez-Cadenas:2023vca}. The detection of~$0\nu\beta\beta$ decay would establish that neutrinos are Majorana fermions and that total lepton number is violated. If light-Majorana-neutrino exchange dominates this process, it also directly probes the absolute neutrino mass scale. The most stringent current constraint is reported by the KamLAND-Zen collaboration, which gives an upper limit range of 0.028--0.122~eV for $m_{\beta \beta}$, depending on the nuclear matrix element used~\cite{KamLAND-Zen:2024eml}. This corresponds to a 90\%~C.L. lower limit on the $0\nu\beta\beta$ decay half-life in~${}^{136}\text{Xe}$ of~$T^{0\nu}_{1/2}>3.8\times10^{26}$~yr. 
An updated overview of various neutrino measurements is provided in~\cite{Nu2024}. 

The usefulness of simple mass relations in particle physics is illustrated by several well-known examples. We recall three such cases here. The first two are the Gatto--Sartori--Tonin and Fritzsch relations~\cite{Gatto:1968ss,Fritzsch:1977za}, which connect the Cabibbo angle to a ratio of down-type quark masses, often expressed as $\sin{\theta_C}\simeq\sqrt{m_d/m_s}$ and~$\tan{\theta_C}\simeq\sqrt{m_d/m_s}$, respectively. The third is the well-known relation among charged-lepton masses found by Koide~\cite{Koide:1982si,Koide:1982ax,Koide:1983qe}, which also takes the form of a simple ratio. When it was proposed in the early 1980's, the Koide relation predicted the tau-lepton mass with high precision. For pole masses,  the formula continues to hold at the~$\sim0.001\%$ level; for running masses, the deviation can increase to~$\sim0.2\%$~\cite{Xing:2006vk}.  These relations have naturally motivated extensions to other fermions, including neutrinos~\cite{Fritzsch:2009sm,Koide:2006dn,Xing:2006vk,Li:2005rp,Cao:2012un,Rodejohann:2011jj,Gao:2015xnv,Rodejohann:2012jz,Hernandez:2013vya,Roy:2020vtm,Mohapatra:2006gs}.  

The purpose of this work is to highlight the possible existence of a simple algebraic relation between~$\Delta m^{2}_{21}$ and $\Delta m^{2}_{31}$, as well as to investigate its implications.

\section{Empirical test}
In particle physics, the ratios of observable quantities—such as scattering cross-sections or magnetic moments—as well as the underlying amplitudes, frequently yield clean fractions or simple square roots like~\(\sqrt{2}\), \(\sqrt{3}\), or \(\sqrt{6}\)~\cite{halzen1984quarks}. Far from being accidental, these geometric and algebraic factors are rooted in the underlying gauge and global symmetries (such as isospin~\(SU(2)\) or flavor \(SU(3)\)) that govern particle interactions. When calculating transition amplitudes across different charge or flavor channels, the complex, non-perturbative dynamical components often cancel out entirely. As a result, the ratio of these  quantities reduces directly to the ratio of their overall normalization factors.

Let us define a simple ratio as a function of the solar and atmospheric neutrino mass-squared differences:
\begin{equation}
\label{eqref:main}
\lambda\left(\Delta m^2_{21},\Delta m^2_{31}\right)=\dfrac{\sqrt{\Delta m^2_{31}}+\sqrt{\Delta m^2_{21}}}{\sqrt{\Delta m^2_{31}}-\sqrt{\Delta m^2_{21}}}.
\end{equation}
Substituting the corresponding best-fit values  from~\cite{deSalas:2020pgw}  into~\eqref{eqref:main} yields 
 \begin{equation}
 \label{eqref:m255}
\lambda\left(7.50\times 10^{-5}~\text{eV}^2, \,2.55\times 10^{-3}~\text{eV}^2\right)=1.4140,\nonumber
\end{equation}
reported to five significant digits.
Remarkably, this is a close approximation to \mbox{$\sqrt{2}\approx1.41421$}.
Parameters within the reported uncertainties can bring the ratio even closer to~$\sqrt{2}$. For example, changing only $\Delta m^2_{21}$ to $7.5065\times 10^{-5}$~eV$^2$—a value within~$0.03\sigma$ of the best fit\footnote{The value $\Delta m^2_{21}=7.5065\times 10^{-5}$~eV$^2$ is likewise very close to the inaugural JUNO central result (within  $0.06\sigma$)~\cite{JUNO:2025gmd}.}—leads to a result approximating~$\sqrt{2}$ to seven decimal places.
We also evaluated this ratio at the best-fit points reported in other major global analyses~\citep{Capozzi:2017ipn,Capozzi:2025wyn,Esteban:2024eli,Esteban:2026phq,Capozzi:2025ovi}. Table~\ref{tab:tab1} demonstrates that all results remain close to~$\sqrt{2}$, with a maximum deviation of~0.27\%. 
\begin{table}[h!]
 \begin{center}
      \caption{Ratio~\eqref{eqref:main} evaluated at several best-fit points for $\Delta m^2_{21}$ and $\Delta m^2_{31}$. A common convention, $\Delta m^2_{31}=m_3^2-m_1^2$, is used; atmospheric splittings from~\cite{Capozzi:2017ipn,Capozzi:2025wyn} are converted to this convention. For the NuFIT entries, $+\mathrm{ATM}$ and $-\mathrm{ATM}$ denote fits with and without the additional atmospheric-neutrino likelihoods, respectively. The results are reported to five significant digits.}
\setlength\extrarowheight{3pt}
      \begin{tabular}{c c c|c}\hline 
      &&&\\[-1em]
      {[Ref.]}&{ $\dfrac{\Delta m^2_{21}}{10^{-5}~\text{eV}^2}$} & { $\dfrac{\Delta m^2_{31}}{10^{-3}~\text{eV}^2}$}&{ $\lambda(\Delta m^2_{21},\Delta m^2_{31})$}\\[1em]
           &{Best fit}& {  Best fit}& { [This work]} \\
       \hline   
     { {\scriptsize Capozzi {\it et al.}\,\,\,\,\,\,\,\,\,\,\,\,\,\,\,\,\,\,\,\, \cite{Capozzi:2017ipn}}}\,\,\,\,\,   &7.34 &2.522 &1.4114\\
    \hline
     { {\scriptsize Capozzi {\it et al.}\,\,\,\,\,\,\,\,\,\,\,\,\,\cite{Capozzi:2025ovi,Capozzi:2025wyn}}}\,\,\,\,\,  &7.48 &2.532 &1.4151\\
     \hline
     { {\scriptsize NuFIT-6.0~($+\mathrm{ATM}$)\,\,\,\, \cite{Esteban:2024eli}}}\,\,\,\,\, &7.49 &2.513 &1.4173\\
     \hline
     { \scriptsize NuFIT-6.0~($-\mathrm{ATM}$)\,\,\,\, \cite{Esteban:2024eli}}\,\,\,\,\, &7.49 &2.534 &1.4152\\
     \hline
     { {\scriptsize de Salas {\it et al.}\,\,\,\,\,\,\,\,\,\,\,\,\,\,\,\,\,\,\, \cite{deSalas:2020pgw}}}\,\,\,\,\, &7.50 &2.550 &1.4140\\
     \hline
      { {\scriptsize NuFIT-6.1~($+\mathrm{ATM}$)\,\,\,\, \cite{Esteban:2026phq}}}\,\,\,\,\, &7.53 &2.519 &1.4181\\
     \hline
     { {\scriptsize NuFIT-6.1~($-\mathrm{ATM}$)\,\,\,\, \cite{Esteban:2026phq}}}\,\,\,\,\, &7.53 &2.529 &1.4171\\
     \hline
      \end{tabular}
      \label{tab:tab1}
  \end{center}
\end{table}
Inspired by these empirical observations, we  propose to assume that
\begin{equation}
\label{eq:main_r}
\frac{\sqrt{\Delta m^2_{31}}+\sqrt{\Delta m^2_{21}}}{\sqrt{\Delta m^2_{31}}-\sqrt{\Delta m^2_{21}}}=\sqrt{2}.
\end{equation}
Taking the ratio to be exactly~$\sqrt{2}$ gives a simple algebraic constraint linking the solar and atmospheric neutrino mass scales. As shown below, this relation has  alternative forms with analogues in the quark and charged-lepton sectors and continues to be compatible with current experimental uncertainties.

It is also tempting to speculate that the emergence of such a specific geometric factor on the right-hand side is a manifestation of underlying quantum dynamics, such as state normalization conditions or specific lepton mixing angle symmetries. While finding a first-principles theoretical derivation of~\eqref{eq:main_r} constitutes an open challenge, we treat it here on purely phenomenological grounds. 

Combining the highly precise value of $\Delta m^2_{21}$ measured by JUNO~\cite{JUNO:2025gmd} with the \mbox{NuFIT-6.1}~result for~$\Delta m^2_{31}$~\cite{Esteban:2026phq}, we find 
\begin{equation} \label{eq:main_predict22} 
\frac{\sqrt{\Delta m^2_{31}}+\sqrt{\Delta m^2_{21}}}{\sqrt{\Delta m^2_{31}}-\sqrt{\Delta m^2_{21}}} = 1.4161 \pm 0.0045.
\end{equation} 
Assuming uncorrelated Gaussian uncertainties, the central value differs from~$\sqrt{2}$ by approximately~$0.4$~standard deviations. This indicates compatibility rather than evidence;
a decisive validation requires the analysis of further independent data. Future measurements and improved fits will test relation~\eqref{eq:main_r} more stringently.

\section{Implications and interpretation \label{sec:consequences}}
In this section, we investigate the implications of adopting the constraint~\eqref{eq:main_r}.
\subsection{How many independent parameters are there?}
Relation~\eqref{eq:main_r} immediately raises the question of how many neutrino mass-squared differences are independent. In the standard three-neutrino treatment, two independent splittings are required~\cite{Denton:2025jkt}. Specifically, the three standard parameters—$\Delta m^2_{21}$, $\Delta m^2_{31}$, and~$\Delta m^2_{32}$—are subject to the consistency condition \mbox{$\Delta m^2_{31}-\Delta m^2_{21}=\Delta m^2_{32}$}. If   relation~\eqref{eq:main_r} also holds, only one independent mass-squared difference remains. The entire spectrum of splittings  is then fixed by a single  scale. 

In other words, the existence of a novel algebraic link between the solar  and atmospheric mass splittings effectively constrains the configuration of the neutrino mass spectrum. Consequently, it removes a critical degree of freedom from theoretical models; rather than requiring separate experimental inputs for both splittings, any complementary theoretical framework needs to introduce only a single overall scale—such as the mass of the lightest neutrino ($m_{1}$) or the cosmological sum ($\sum m_\nu$)—to completely determine the individual absolute masses. This reduction in free parameters would significantly enhance the predictive power of models aiming to uncover the origin of neutrino masses.

\subsection{Koide-type representation}

Straightforward algebra transforms relation~\eqref{eq:main_r} into
\begin{equation}
\label{eq:ratio34}
\frac{\Delta m^2_{21}+\Delta m^2_{31}}{\left(\sqrt{\Delta m^2_{21}}+\sqrt{\Delta m^2_{31}}\right)^2}=\frac{3}{4}.
\end{equation}
This is likewise consistent with the global fits~\citep{deSalas:2020pgw,Capozzi:2017ipn,Capozzi:2025wyn,Esteban:2024eli,Esteban:2026phq}. For example, with the data from~\citep{deSalas:2020pgw}, the left-hand side is~$0.7501$, close to  the predicted value of~3/4. 
Equation~\eqref{eq:ratio34}~mirrors the structural form of the Koide formula for charged-lepton masses~\cite{Koide:1982si,Koide:1982ax,Koide:1983qe}, 
\begin{equation}
\label{eq:koide}
\frac{m_e+m_{\mu}+m_{\tau}}{\left(\sqrt{m_e}+\sqrt{m_{\mu}}+\sqrt{m_{\tau}}\right)^2}=\frac{2}{3},
\end{equation}
with the neutrino mass-squared differences as the structural components.   
For context, the Koide formula  holds at the $\sim0.001\%$ level for pole masses but can deviate by approximately 0.2\% for running charged-lepton masses at higher energy scales, such as the $Z$~pole~($M_Z$)~\cite{Xing:2006vk}. It is also relevant to note that extensions of Koide’s result to the neutrino sector have been widely investigated~\cite{Koide:2006dn,Xing:2006vk,Li:2005rp,Cao:2012un,Rodejohann:2011jj,Gao:2015xnv}.

\subsection{Fritzsch-type representation}
Relation~\eqref{eq:main_r} can equivalently be represented as
\begin{equation}
\label{eq:Fritzsch}
\sqrt[4]{\frac{\Delta m^2_{21}}{\Delta m^2_{31}}}=\tan\frac{\pi}{8},
\end{equation}
This form is also in good agreement with the recent global fits~\citep{deSalas:2020pgw,Capozzi:2017ipn,Capozzi:2025wyn,Esteban:2024eli,Esteban:2026phq}. Notably, for a massless lightest neutrino,~$m_1=0$, \eqref{eq:Fritzsch}~reduces to
\begin{equation}
\label{eq:Fritzsch1}
\sqrt{\frac{m_2}{m_3}}=\tan\frac{\pi}{8}, 
\end{equation}
which coincides with the Fritzsch relation for fundamental-fermion masses~\cite{Fritzsch:1977za} at~${\theta=\pi/8}$ $(22.5^\circ)$. It is interesting to speculate that this correspondence points toward a vanishing $\nu_1$~mass.
Moreover, because~\mbox{$\pi/8\approx\theta_{23}/2$}~\cite{nufit61}, \eqref{eq:Fritzsch}~ again raises  the question of how many parameters in the neutrino sector are genuinely independent.

\subsection{Algebraic structure of neutrino masses}
Relation~\eqref{eq:main_r} also implies 
\begin{equation}
\label{eq:deltam32i}
\left(\Delta m^2_{31}-\Delta m^2_{21}\right)^2=32\,\Delta m^2_{21}\Delta m^2_{31},
\end{equation}
which can equivalently be written as
\begin{equation}
\label{eq:deltam32}
m^2_3-m^2_2=\sqrt{32\Delta m^2_{21}\Delta m^2_{31}}.
\end{equation}
This explicitly shows that the mass-squared-difference spectrum is fixed by a single splitting scale.
A useful implication of~\eqref{eq:deltam32} is that it constrains the functional forms of $m_2$ and $m_3$ in theoretical models where these masses are expressed in terms of~$\Delta m^2_{21}$ and $\Delta m^2_{31}$ (see, e.g.,~\cite{Fritzsch:2006sm}). Any prediction from such a  model must satisfy~\eqref{eq:deltam32}. Equation~ \eqref{eq:deltam32} may also serve as a heuristic  guide  to the algebraic structure of the neutrino masses. This  will be investigated in more detail elsewhere.

The formulae \eqref{eq:ratio34},
\eqref{eq:Fritzsch}, and \eqref{eq:deltam32} are algebraically equivalent representations of the same relation, not independent on each other. Future neutrino-oscillation data with higher statistics, particularly from JUNO~\cite{JUNO:2022mxj}, together with new global fits, will provide stricter tests of these results.

\subsection{Derivation of $\Delta m^2_{31}$ from $\Delta m^2_{21}$ and vice versa}

Crucially, relation~\eqref{eq:main_r} has  clear practical utility given its current phenomenological status. By adopting it, we obtain an elegant algebraic equation linking~$\Delta m^2_{21}$ and $\Delta m^2_{31}$. Therefore, knowing one of these parameters, we can readily find the other. 

As a concrete application, we can predict the value of $\Delta m^2_{31}$ expected at JUNO by leveraging the experiment's initial measurement of $\Delta m^2_{21}=(7.50\pm0.12)\times10^{-5}~\text{eV}^2$~\cite{JUNO:2025gmd}. 
Solving~\eqref{eq:main_r} for $\Delta m^2_{31}$ yields

\begin{equation}
\label{eq:solution_m31}
\Delta m^2_{31}=\left(\sqrt{8}+\sqrt{9}\right)^2\Delta m^2_{21}.
\end{equation}
Substituting the JUNO result into~\eqref{eq:solution_m31}, we arrive at

\begin{equation}
\label{eq:solution m31 num}
\Delta m^2_{31}=(2.548\pm0.041)\times10^{-3}~\text{eV}^2.
\end{equation}
This outcome exhibits good consistency with neutrino oscillation data~\citep{Capozzi:2017ipn,Capozzi:2025wyn,Esteban:2024eli,Esteban:2026phq}, while  with roughly twice the uncertainty of the global fit frameworks, such as NuFIT-6.1. However, solution~\eqref{eq:solution_m31} naturally allows the numerical prediction to be refined as new data become available.
The JUNO experiment aims to determine~$\Delta m^2_{21}$ and $\Delta m^2_{31}$ at the sub-percent level, thereby providing a direct test of our results. 

Conversely, relation~\eqref{eq:main_r} enables us to  derive the solar mass splitting from $\Delta m^2_{31}$:

\begin{equation}
\label{eq:solution_m21}
\Delta m^2_{21}=\left(\sqrt{9}-\sqrt{8}\right)^2\Delta m^2_{31}.
\end{equation}
Using the recent NuFIT-6.1 value of $\Delta m^2_{31}=(2.529\pm0.021)\times10^{-3}~\text{eV}^2$~\cite{Esteban:2026phq}, we find

\begin{equation}
\label{eq:solution m21 num}
\Delta m^2_{21}=(7.445\pm0.062)\times10^{-5}~\text{eV}^2.
\end{equation} 
A nearly identical prediction is obtained if one instead employs the data from~\cite{Capozzi:2025wyn}. Remarkably, \eqref{eq:solution m21 num}~constitutes a quite stringent constraint on the solar mass splitting. It is significantly tighter than the first JUNO result~\cite{JUNO:2025gmd} and  the global averages~\citep{Capozzi:2017ipn,Capozzi:2025wyn,Esteban:2024eli,Esteban:2026phq,Capozzi:2025ovi}. For comparison, Capozzi et al. report $\Delta m^2_{21}=\left(7.48\pm0.10\right)\times10^{-5}~\text{eV}^2$~\citep{Capozzi:2025ovi}.



\section{Conclusions \label{sec:conclusion}}

Determining the absolute neutrino mass scale remains a central challenge in particle physics. Establishing precise mathematical connections among the neutrino masses could significantly facilitate this task and shed additional light on the underlying mass-generation mechanism. In this work, we have highlighted the possible existence of an algebraic relation between the mass-squared differences; its simple ratio format agrees remarkably well with the best-fit values of recent global analyses of neutrino oscillation data, such as~NuFIT{$\text{-}6.0/6.1$}.

Adopting this relation, a single splitting scale determines the entire mass-squared-difference spectrum. This raises the question of how many oscillation parameters are genuinely independent.  Notably, assuming a vanishing lightest neutrino mass ($m_1=0$) directly reduces our ansatz to the elegant Fritzsch relation. The ansatz also yields an expression mirroring the structural form of the Koide formula~\cite{Koide:1982si,Koide:1982ax,Koide:1983qe}, complementing existing neutrino-mass extensions of the latter~\cite{Koide:2006dn,Xing:2006vk,Li:2005rp,Cao:2012un,Rodejohann:2011jj,Gao:2015xnv}. It further implies that the product $\Delta m^2_{21}\Delta m^2_{31}$ is fixed by the difference $\Delta m_{32}^2 = m_3^2 - m_2^2$, thereby constraining the allowed functional forms of $m_2$ and $m_3$. Throughout the analysis, we assumed normal ordering ($\Delta m^2_{31}>0$). An extension to inverted ordering would require a separately specified positive convention for the atmospheric splitting.

The appearance of specific numbers such as $\sqrt{2}$ in~\eqref{eq:main_r} may reflect more than a numerical coincidence. Such factors can arise from state normalization conditions or geometric mixing structures. Deriving the relation within a quantum theoretical framework would provide a rigorous foundation for our phenomenological approach.

The proposed relation can also serve as a useful reference
ansatz when the measured value differs slightly
from the idealized~$\sqrt{2}$~baseline. As seen from Table~\ref{tab:tab1} and~\eqref{eq:main_predict22}, this ratio remains stable to at least two decimal places ($\approx1.41$), allowing small deviations to be treated  perturbatively around~\eqref{eq:main_r}. Crucially, the best fit values of the oscillation parameters approximate~$\sqrt{2}$ with a maximum relative error of only $0.27\%$. This represents excellent agreement in the context of neutrino physics, especially considering that the global fits were derived completely independently of our results. Adopting~\eqref{eq:main_r} as an exact equation for  the oscillation parameters and relying on recent global fits, we have placed a tight constraint on the solar mass splitting $\Delta m_{21}^{2}$. Likewise, we have numerically determined the atmospheric mass spliting~$\Delta m_{31}^{2}$ by leveraging the first JUNO data.

Using the JUNO and \mbox{NuFIT-6.1} values quoted above and independent Gaussian error
propagation, the measured ratio differs from~$\sqrt{2}$ by approximately~$0.4$ standard deviations. Since the listed global analyses share common data and assumptions, this consistency, while suggestive, does not yet constitute statistical evidence for a novel regularity among the oscillation parameters. A future test should treat~\eqref{eq:main_r} as a fixed prediction and evaluate it using an independent data release and the full joint likelihood. 
Upcoming experimental results, particularly from JUNO~\cite{JUNO:2022mxj}, will provide a stringent probe of this hypothesis and its consequences. JUNO's first~59.1~days of data already yielded a leading measurement of~$\Delta m^2_{21}$~\cite{JUNO:2025gmd}, illustrating its potential in the field. With six years of data, this experiment is expected to reach sub-percent precision~($\lesssim 0.3\%$) for both parameters: $\Delta m_{21}^{2}$ and $\Delta m_{31}^{2}$.


\acknowledgments
The author thanks J.~F.~Beacom and T.~Schwetz for useful correspondence.
\vskip 1cm
\noindent Note added: After this manuscript was finished, the JUNO Collaboration reported the most precise measurements to date of both neutrino mass splittings~\cite{JUNO:2026tap}, which further support relation~\eqref{eq:main_r} and its consequences. A detailed comparison of the relation with these new experimental data will be performed elsewhere.

\bibliographystyle{JHEP}
\bibliography{refs}

\providecommand{\href}[2]{#2}\begingroup\raggedright\begin{thebibliography}{10}

\bibitem{ParticleDataGroup:2024cfk}
{\scshape Particle Data Group} collaboration, \emph{{Review of particle
  physics}}, \href{https://doi.org/10.1103/PhysRevD.110.030001}{\emph{Phys.
  Rev. D} {\bfseries 110} (2024) 030001}.

\bibitem{SNO:2002tuh}
{\scshape SNO} collaboration, \emph{{Direct evidence for neutrino flavor
  transformation from neutral current interactions in the Sudbury Neutrino
  Observatory}},
  \href{https://doi.org/10.1103/PhysRevLett.89.011301}{\emph{Phys. Rev. Lett.}
  {\bfseries 89} (2002) 011301}
  [\href{https://arxiv.org/abs/nucl-ex/0204008}{{\ttfamily nucl-ex/0204008}}].

\bibitem{Super-Kamiokande:2001ljr}
{\scshape Super-Kamiokande} collaboration, \emph{{Solar B-8 and hep neutrino
  measurements from 1258 days of Super-Kamiokande data}},
  \href{https://doi.org/10.1103/PhysRevLett.86.5651}{\emph{Phys. Rev. Lett.}
  {\bfseries 86} (2001) 5651}
  [\href{https://arxiv.org/abs/hep-ex/0103032}{{\ttfamily hep-ex/0103032}}].

\bibitem{Super-Kamiokande:1998kpq}
{\scshape Super-Kamiokande} collaboration, \emph{{Evidence for oscillation of
  atmospheric neutrinos}},
  \href{https://doi.org/10.1103/PhysRevLett.81.1562}{\emph{Phys. Rev. Lett.}
  {\bfseries 81} (1998) 1562}
  [\href{https://arxiv.org/abs/hep-ex/9807003}{{\ttfamily hep-ex/9807003}}].

\bibitem{IceCube:2014flw}
{\scshape IceCube} collaboration, \emph{{Determining neutrino oscillation
  parameters from atmospheric muon neutrino disappearance with three years of
  IceCube DeepCore data}},
  \href{https://doi.org/10.1103/PhysRevD.91.072004}{\emph{Phys. Rev. D}
  {\bfseries 91} (2015) 072004}
  [\href{https://arxiv.org/abs/1410.7227}{{\ttfamily 1410.7227}}].

\bibitem{K2K:2004iot}
{\scshape K2K} collaboration, \emph{{Evidence for muon neutrino oscillation in
  an accelerator-based experiment}},
  \href{https://doi.org/10.1103/PhysRevLett.94.081802}{\emph{Phys. Rev. Lett.}
  {\bfseries 94} (2005) 081802}
  [\href{https://arxiv.org/abs/hep-ex/0411038}{{\ttfamily hep-ex/0411038}}].

\bibitem{MINOS:2008kxu}
{\scshape MINOS} collaboration, \emph{{Measurement of Neutrino Oscillations
  with the MINOS Detectors in the NuMI Beam}},
  \href{https://doi.org/10.1103/PhysRevLett.101.131802}{\emph{Phys. Rev. Lett.}
  {\bfseries 101} (2008) 131802}
  [\href{https://arxiv.org/abs/0806.2237}{{\ttfamily 0806.2237}}].

\bibitem{T2K:2011ypd}
{\scshape T2K} collaboration, \emph{{Indication of Electron Neutrino Appearance
  from an Accelerator-produced Off-axis Muon Neutrino Beam}},
  \href{https://doi.org/10.1103/PhysRevLett.107.041801}{\emph{Phys. Rev. Lett.}
  {\bfseries 107} (2011) 041801}
  [\href{https://arxiv.org/abs/1106.2822}{{\ttfamily 1106.2822}}].

\bibitem{OPERA:2015wbl}
{\scshape OPERA} collaboration, \emph{{Discovery of $\tau$ Neutrino Appearance
  in the CNGS Neutrino Beam with the OPERA Experiment}},
  \href{https://doi.org/10.1103/PhysRevLett.115.121802}{\emph{Phys. Rev. Lett.}
  {\bfseries 115} (2015) 121802}
  [\href{https://arxiv.org/abs/1507.01417}{{\ttfamily 1507.01417}}].

\bibitem{KamLAND:2002uet}
{\scshape KamLAND} collaboration, \emph{{First results from KamLAND: Evidence
  for reactor anti-neutrino disappearance}},
  \href{https://doi.org/10.1103/PhysRevLett.90.021802}{\emph{Phys. Rev. Lett.}
  {\bfseries 90} (2003) 021802}
  [\href{https://arxiv.org/abs/hep-ex/0212021}{{\ttfamily hep-ex/0212021}}].

\bibitem{DoubleChooz:2011ymz}
{\scshape Double Chooz} collaboration, \emph{{Indication of Reactor
  $\bar{\nu}_e$ Disappearance in the Double Chooz Experiment}},
  \href{https://doi.org/10.1103/PhysRevLett.108.131801}{\emph{Phys. Rev. Lett.}
  {\bfseries 108} (2012) 131801}
  [\href{https://arxiv.org/abs/1112.6353}{{\ttfamily 1112.6353}}].

\bibitem{DayaBay:2012fng}
{\scshape Daya Bay} collaboration, \emph{{Observation of electron-antineutrino
  disappearance at Daya Bay}},
  \href{https://doi.org/10.1103/PhysRevLett.108.171803}{\emph{Phys. Rev. Lett.}
  {\bfseries 108} (2012) 171803}
  [\href{https://arxiv.org/abs/1203.1669}{{\ttfamily 1203.1669}}].

\bibitem{RENO:2012mkc}
{\scshape RENO} collaboration, \emph{{Observation of Reactor Electron
  Antineutrino Disappearance in the RENO Experiment}},
  \href{https://doi.org/10.1103/PhysRevLett.108.191802}{\emph{Phys. Rev. Lett.}
  {\bfseries 108} (2012) 191802}
  [\href{https://arxiv.org/abs/1204.0626}{{\ttfamily 1204.0626}}].

\bibitem{T2K:2023smv}
{\scshape T2K} collaboration, \emph{{Measurements of neutrino oscillation
  parameters from the T2K experiment using $3.6\times 10^{21}$ protons on
  target}}, \href{https://doi.org/10.1140/epjc/s10052-023-11819-x}{\emph{Eur.
  Phys. J. C} {\bfseries 83} (2023) 782}
  [\href{https://arxiv.org/abs/2303.03222}{{\ttfamily 2303.03222}}].

\bibitem{T2K:2025yoy}
{\scshape T2K} collaboration, \emph{{Results from the T2K Experiment on
  Neutrino Mixing Including a New Far Detector {\ensuremath{\mu}}-like
  Sample}}, \href{https://doi.org/10.1103/gh5j-5cwv}{\emph{Phys. Rev. Lett.}
  {\bfseries 135} (2025) 261801}
  [\href{https://arxiv.org/abs/2506.05889}{{\ttfamily 2506.05889}}].

\bibitem{NOvA:2025tmb}
{\scshape NOvA} collaboration, \emph{{Precision Measurement of Neutrino
  Oscillation Parameters with 10 Years of Data from the NOvA Experiment}},
  \href{https://doi.org/10.1103/x53y-2b86}{\emph{Phys. Rev. Lett.} {\bfseries
  136} (2026) 011802} [\href{https://arxiv.org/abs/2509.04361}{{\ttfamily
  2509.04361}}].

\bibitem{Esteban:2026phq}
I.~Esteban, M.C.~Gonzalez-Garcia, M.~Maltoni, I.~Martinez-Soler, J.P.~Pinheiro
  and T.~Schwetz, \emph{{Lessons from the first JUNO results}},
  \href{https://doi.org/10.1007/JHEP04(2026)089}{\emph{JHEP} {\bfseries 04}
  (2026) 089} [\href{https://arxiv.org/abs/2601.09791}{{\ttfamily
  2601.09791}}].

\bibitem{T2K:2025wet}
{\scshape T2K, NOvA} collaboration, \emph{{Joint neutrino oscillation analysis
  from the T2K and NOvA experiments}},
  \href{https://doi.org/10.1038/s41586-025-09599-3}{\emph{Nature} {\bfseries
  646} (2025) 818} [\href{https://arxiv.org/abs/2510.19888}{{\ttfamily
  2510.19888}}].

\bibitem{JUNO:2022mxj}
{\scshape JUNO} collaboration, \emph{{Sub-percent precision measurement of
  neutrino oscillation parameters with JUNO}},
  \href{https://doi.org/10.1088/1674-1137/ac8bc9}{\emph{Chin. Phys. C}
  {\bfseries 46} (2022) 123001}
  [\href{https://arxiv.org/abs/2204.13249}{{\ttfamily 2204.13249}}].

\bibitem{JUNO:2025gmd}
{\scshape JUNO} collaboration, \emph{{Measurement of reactor neutrino
  oscillation with the first JUNO data}},
  \href{https://doi.org/10.1038/s41586-026-10538-z}{\emph{Nature} {\bfseries
  654} (2026) 343} [\href{https://arxiv.org/abs/2511.14593}{{\ttfamily
  2511.14593}}].

\bibitem{DUNE:2020lwj}
{\scshape DUNE} collaboration, \emph{{Deep Underground Neutrino Experiment
  (DUNE), Far Detector Technical Design Report, Volume I Introduction to
  DUNE}}, \href{https://doi.org/10.1088/1748-0221/15/08/T08008}{\emph{JINST}
  {\bfseries 15} (2020) T08008}
  [\href{https://arxiv.org/abs/2002.02967}{{\ttfamily 2002.02967}}].

\bibitem{Hyper-Kamiokande:2018ofw}
{\scshape Hyper-Kamiokande} collaboration, \emph{{Hyper-Kamiokande Design
  Report}},  \href{https://arxiv.org/abs/1805.04163}{{\ttfamily 1805.04163}}.

\bibitem{Denton:2025jkt}
P.B.~Denton, \emph{{Neutrino Oscillations in the Three Flavor Paradigm}},  in
  \emph{{Encyclopedia of Particle Physics: Volume 3}}, C.S.~Fischer, A.~Pich,
  A.B.~Balantekin and V.~Sanz, eds., vol.~3, pp.~194--217, Elsevier (2026)
  [\href{https://arxiv.org/abs/2501.08374}{{\ttfamily 2501.08374}}].

\bibitem{Khan:2019doq}
A.N.~Khan, H.~Nunokawa and S.J.~Parke, \emph{{Why matter effects matter for
  JUNO}}, \href{https://doi.org/10.1016/j.physletb.2020.135354}{\emph{Phys.
  Lett. B} {\bfseries 803} (2020) 135354}
  [\href{https://arxiv.org/abs/1910.12900}{{\ttfamily 1910.12900}}].

\bibitem{Drexlin:2026zam}
G.~Drexlin and C.~Weinheimer, \emph{{KATRIN experiment}},
  \href{https://arxiv.org/abs/2601.00248}{{\ttfamily 2601.00248}}.

\bibitem{Dolgov:2002wy}
A.D.~Dolgov, \emph{{Neutrinos in cosmology}},
  \href{https://doi.org/10.1016/S0370-1573(02)00139-4}{\emph{Phys. Rept.}
  {\bfseries 370} (2002) 333}
  [\href{https://arxiv.org/abs/hep-ph/0202122}{{\ttfamily hep-ph/0202122}}].

\bibitem{KATRIN:2024cdt}
{\scshape KATRIN} collaboration, \emph{{Direct neutrino-mass measurement based
  on 259 days of KATRIN data}},
  \href{https://doi.org/10.1126/science.adq9592}{\emph{Science} {\bfseries 388}
  (2025) adq9592} [\href{https://arxiv.org/abs/2406.13516}{{\ttfamily
  2406.13516}}].

\bibitem{DESI:2024mwx}
{\scshape DESI} collaboration, \emph{{DESI 2024 VI: cosmological constraints
  from the measurements of baryon acoustic oscillations}},
  \href{https://doi.org/10.1088/1475-7516/2025/02/021}{\emph{JCAP} {\bfseries
  02} (2025) 021} [\href{https://arxiv.org/abs/2404.03002}{{\ttfamily
  2404.03002}}].

\bibitem{Naredo-Tuero:2024sgf}
D.~Naredo-Tuero, M.~Escudero, E.~Fern{\'a}ndez-Mart{\'\i}nez, X.~Marcano and
  V.~Poulin, \emph{{Critical look at the cosmological neutrino mass bound}},
  \href{https://doi.org/10.1103/PhysRevD.110.123537}{\emph{Phys. Rev. D}
  {\bfseries 110} (2024) 123537}
  [\href{https://arxiv.org/abs/2407.13831}{{\ttfamily 2407.13831}}].

\bibitem{Gorbunov:2026sly}
D.~Gorbunov and N.~Nedelko, \emph{{Changing constraints on
  {\ensuremath{\sum}}m{\ensuremath{\nu}} from SPT-3G 2018 to D1 in the context
  of DESI DR2}},
  \href{https://doi.org/10.1016/j.physletb.2026.140659}{\emph{Phys. Lett. B}
  {\bfseries 879} (2026) 140659}
  [\href{https://arxiv.org/abs/2601.16277}{{\ttfamily 2601.16277}}].

\bibitem{DellOro:2016tmg}
S.~Dell'Oro, S.~Marcocci, M.~Viel and F.~Vissani, \emph{{Neutrinoless double
  beta decay: 2015 review}},
  \href{https://doi.org/10.1155/2016/2162659}{\emph{Adv. High Energy Phys.}
  {\bfseries 2016} (2016) 2162659}
  [\href{https://arxiv.org/abs/1601.07512}{{\ttfamily 1601.07512}}].

\bibitem{Gomez-Cadenas:2023vca}
J.J.~G{\'o}mez-Cadenas, J.~Mart{\'\i}n-Albo, J.~Men{\'e}ndez, M.~Mezzetto,
  F.~Monrabal and M.~Sorel, \emph{{The search for neutrinoless double-beta
  decay}}, \href{https://doi.org/10.1007/s40766-023-00049-2}{\emph{Riv. Nuovo
  Cim.} {\bfseries 46} (2023) 619}.

\bibitem{KamLAND-Zen:2024eml}
{\scshape KamLAND-Zen} collaboration, \emph{{Search for Majorana Neutrinos with
  the Complete KamLAND-Zen Dataset}},
  \href{https://doi.org/10.1103/jkf6-48j8}{\emph{Phys. Rev. Lett.} {\bfseries
  135} (2025) 262501} [\href{https://arxiv.org/abs/2406.11438}{{\ttfamily
  2406.11438}}].

\bibitem{Nu2024}
\emph{{ Neutrino 2024, {\em XXXI International Conference on Neutrino Physics
  and Astrophysics}}}, (Milan, Italy), June.

\bibitem{Gatto:1968ss}
R.~Gatto, G.~Sartori and M.~Tonin, \emph{{Weak Selfmasses, Cabibbo Angle, and
  Broken SU(2) x SU(2)}},
  \href{https://doi.org/10.1016/0370-2693(68)90150-0}{\emph{Phys. Lett. B}
  {\bfseries 28} (1968) 128}.

\bibitem{Fritzsch:1977za}
H.~Fritzsch, \emph{{Calculating the Cabibbo Angle}},
  \href{https://doi.org/10.1016/0370-2693(77)90408-7}{\emph{Phys. Lett. B}
  {\bfseries 70} (1977) 436}.

\bibitem{Koide:1982si}
Y.~Koide, \emph{{Fermion - Boson Two-body Model of Quarks and Leptons and
  Cabibbo Mixing}}, \href{https://doi.org/10.1007/BF02817096}{\emph{Lett. Nuovo
  Cim.} {\bfseries 34} (1982) 201}.

\bibitem{Koide:1982ax}
Y.~Koide, \emph{{A Fermion - Boson Composite Model of Quarks and Leptons}},
  \href{https://doi.org/10.1016/0370-2693(83)90644-5}{\emph{Phys. Lett. B}
  {\bfseries 120} (1983) 161}.

\bibitem{Koide:1983qe}
Y.~Koide, \emph{{A New View of Quark and Lepton Mass Hierarchy}},
  \href{https://doi.org/10.1103/PhysRevD.28.252}{\emph{Phys. Rev. D} {\bfseries
  28} (1983) 252}.

\bibitem{Xing:2006vk}
Z.-z.~Xing and H.~Zhang, \emph{{On the Koide-like relations for the running
  masses of charged leptons, neutrinos and quarks}},
  \href{https://doi.org/10.1016/j.physletb.2006.02.051}{\emph{Phys. Lett. B}
  {\bfseries 635} (2006) 107}
  [\href{https://arxiv.org/abs/hep-ph/0602134}{{\ttfamily hep-ph/0602134}}].

\bibitem{Fritzsch:2009sm}
H.~Fritzsch and Z.-z.~Xing, \emph{{Relating the neutrino mixing angles to a
  lepton mass hierarchy}},
  \href{https://doi.org/10.1016/j.physletb.2009.11.018}{\emph{Phys. Lett. B}
  {\bfseries 682} (2009) 220}
  [\href{https://arxiv.org/abs/0911.1857}{{\ttfamily 0911.1857}}].

\bibitem{Koide:2006dn}
Y.~Koide, \emph{{Tribimaximal Neutrino Mixing and a Relation Between Neutrino-
  and Charged Lepton-Mass Spectra}},
  \href{https://doi.org/10.1088/0954-3899/34/7/006}{\emph{J. Phys. G}
  {\bfseries 34} (2007) 1653}
  [\href{https://arxiv.org/abs/hep-ph/0605074}{{\ttfamily hep-ph/0605074}}].

\bibitem{Li:2005rp}
N.~Li and B.-Q.~Ma, \emph{{Estimate of neutrino masses from Koide's relation}},
  \href{https://doi.org/10.1016/j.physletb.2005.01.066}{\emph{Phys. Lett. B}
  {\bfseries 609} (2005) 309}
  [\href{https://arxiv.org/abs/hep-ph/0505028}{{\ttfamily hep-ph/0505028}}].

\bibitem{Cao:2012un}
F.-G.~Cao, \emph{{Neutrino masses from lepton and quark mass relations and
  neutrino oscillations}},
  \href{https://doi.org/10.1103/PhysRevD.85.113003}{\emph{Phys. Rev. D}
  {\bfseries 85} (2012) 113003}
  [\href{https://arxiv.org/abs/1205.4068}{{\ttfamily 1205.4068}}].

\bibitem{Rodejohann:2011jj}
W.~Rodejohann and H.~Zhang, \emph{{Extension of an empirical charged lepton
  mass relation to the neutrino sector}},
  \href{https://doi.org/10.1016/j.physletb.2011.03.007}{\emph{Phys. Lett. B}
  {\bfseries 698} (2011) 152}
  [\href{https://arxiv.org/abs/1101.5525}{{\ttfamily 1101.5525}}].

\bibitem{Gao:2015xnv}
G.-H.~Gao and N.~Li, \emph{{Explorations of two empirical formulas for fermion
  masses}}, \href{https://doi.org/10.1140/epjc/s10052-016-3990-3}{\emph{Eur.
  Phys. J. C} {\bfseries 76} (2016) 140}
  [\href{https://arxiv.org/abs/1512.06349}{{\ttfamily 1512.06349}}].

\bibitem{Rodejohann:2012jz}
W.~Rodejohann, M.~Tanimoto and A.~Watanabe, \emph{{Relating large $U_{e3}$ to
  the ratio of neutrino mass-squared differences}},
  \href{https://doi.org/10.1016/j.physletb.2012.03.037}{\emph{Phys. Lett. B}
  {\bfseries 710} (2012) 636}
  [\href{https://arxiv.org/abs/1201.4936}{{\ttfamily 1201.4936}}].

\bibitem{Hernandez:2013vya}
D.~Hernandez and A.Y.~Smirnov, \emph{{Relating neutrino masses and mixings by
  discrete symmetries}},
  \href{https://doi.org/10.1103/PhysRevD.88.093007}{\emph{Phys. Rev. D}
  {\bfseries 88} (2013) 093007}
  [\href{https://arxiv.org/abs/1304.7738}{{\ttfamily 1304.7738}}].

\bibitem{Roy:2020vtm}
S.~Roy, K.~Sashikanta~Singh and J.~Borah, \emph{{Revamped Bi-Large neutrino
  mixing with Gatto-Sartori-Tonin like relation}},
  \href{https://doi.org/10.1016/j.nuclphysb.2020.115204}{\emph{Nucl. Phys. B}
  {\bfseries 960} (2020) 115204}
  [\href{https://arxiv.org/abs/2001.07401}{{\ttfamily 2001.07401}}].

\bibitem{Mohapatra:2006gs}
R.N.~Mohapatra and A.Y.~Smirnov, \emph{{Neutrino Mass and New Physics}},
  \href{https://doi.org/10.1146/annurev.nucl.56.080805.140534}{\emph{Ann. Rev.
  Nucl. Part. Sci.} {\bfseries 56} (2006) 569}
  [\href{https://arxiv.org/abs/hep-ph/0603118}{{\ttfamily hep-ph/0603118}}].

\bibitem{halzen1984quarks}
F.~Halzen and A.D.~Martin, \emph{Quarks and Leptons: An Introductory Course in
  Modern Particle Physics}, John Wiley \& Sons, New York (1984).

\bibitem{deSalas:2020pgw}
P.F.~de~Salas, D.V.~Forero, S.~Gariazzo, P.~Mart{\'\i}nez-Mirav{\'e}, O.~Mena,
  C.A.~Ternes et~al., \emph{{2020 Global reassessment of the neutrino
  oscillation picture}},
  \href{https://doi.org/10.1007/JHEP02(2021)071}{\emph{JHEP} {\bfseries 02}
  (2021) 071} [\href{https://arxiv.org/abs/2006.11237}{{\ttfamily
  2006.11237}}].

\bibitem{Capozzi:2017ipn}
F.~Capozzi, E.~Di~Valentino, E.~Lisi, A.~Marrone, A.~Melchiorri and A.~Palazzo,
  \emph{{Addendum to ``Global constraints on absolute neutrino masses and their
  ordering"}}, \href{https://doi.org/10.1103/PhysRevD.101.116013}{\emph{Phys.
  Rev. D} {\bfseries 101} (2020) 116013}.

\bibitem{Capozzi:2025wyn}
F.~Capozzi, W.~Giar{\`e}, E.~Lisi, A.~Marrone, A.~Melchiorri and A.~Palazzo,
  \emph{{Neutrino masses and mixing: Entering the era of subpercent
  precision}}, \href{https://doi.org/10.1103/PhysRevD.111.093006}{\emph{Phys.
  Rev. D} {\bfseries 111} (2025) 093006}
  [\href{https://arxiv.org/abs/2503.07752}{{\ttfamily 2503.07752}}].

\bibitem{Esteban:2024eli}
I.~Esteban, M.C.~Gonzalez-Garcia, M.~Maltoni, I.~Martinez-Soler, J.P.~Pinheiro
  and T.~Schwetz, \emph{{NuFit-6.0: updated global analysis of three-flavor
  neutrino oscillations}},
  \href{https://doi.org/10.1007/JHEP12(2024)216}{\emph{JHEP} {\bfseries 12}
  (2024) 216} [\href{https://arxiv.org/abs/2410.05380}{{\ttfamily
  2410.05380}}].

\bibitem{Capozzi:2025ovi}
F.~Capozzi, E.~Lisi, F.~Marcone, A.~Marrone and A.~Palazzo, \emph{{Updated
  bounds on the (1, 2) neutrino oscillation parameters after first JUNO
  results}}, \href{https://doi.org/10.1103/cxqw-1bty}{\emph{Phys. Rev. D}
  {\bfseries 114} (2026) 016026}
  [\href{https://arxiv.org/abs/2511.21650}{{\ttfamily 2511.21650}}].

\bibitem{nufit61}
I.~Esteban, M.C.~Gonzalez-Garcia, M.~Maltoni, I.~Martinez-Soler, J.P.~Pinheiro
  and T.~Schwetz, ``{NuFIT 6.1}.'' \url{http://www.nu-fit.org}, 2025.

\bibitem{Fritzsch:2006sm}
H.~Fritzsch and Z.-z.~Xing, \emph{{Lepton mass hierarchy and neutrino mixing}},
  \href{https://doi.org/10.1016/j.physletb.2006.02.028}{\emph{Phys. Lett. B}
  {\bfseries 634} (2006) 514}
  [\href{https://arxiv.org/abs/hep-ph/0601104}{{\ttfamily hep-ph/0601104}}].

\bibitem{JUNO:2026tap}
{\scshape JUNO} collaboration, \emph{{Measurement of Neutrino Oscillation
  Parameters at JUNO and Indication of the Neutrino Mass Ordering}},
  \href{https://arxiv.org/abs/2609.37895}{{\ttfamily 2609.37895}}.

\end{thebibliography}\endgroup
\end{document}